# A Recommendation System-based Framework for Enhancing Human-Machine Collaboration in Industrial Timetabling Rescheduling: Application in Preventive Maintenance

Kévin Ducharlet[1], Liwen Zhang[1], Sara Maqrot[1], and Houssem Saidi[1]

[1] Berger-Levrault, Labège, France
`[firstname.lastname]@berger-levrault.com`

**Abstract.** Industrial timetabling is a critical task for decision-makers across various sectors to ensure efficient system operation. In real-world settings, it remains challenging because unexpected events often disrupt execution. When such events arise, effective rescheduling and collaboration between humans and machines becomes essential. This paper presents a recommendation system-based framework for handling rescheduling challenges, built on Timefold, a powerful AI-driven planning engine. Our experimental study evaluates nine instances inspired by a real-world preventive maintenance use case, aiming to identify the heuristic that best balances solution quality and computing time to support near-optimal decision-making when rescheduling is required due to unexpected events during operational days. Finally, we illustrate the complete process of our recommendation system through a simple use case.

## 1 Introduction

Industrial timetabling is a central task for decision-makers across various sectors, ensuring the efficient operation of a given system. As the definition of timetabling by [1]: "*Timetabling is the allocation, subject to constraints, of given resources to objects being placed in space time, in such a way as to satisfy as nearly as possible a set of desirable objectives*". Solving such problems enables decision-makers to allocate resources efficiently, ensuring that the right resources are assigned to the appropriate individuals within the specified time windows respecting various constraints set by the enterprise. In the decision-making process, decision-makers must clearly define their objectives, specifying the goals to be achieved or the risks to be avoided. In general, Operations Research (OR) is the discipline of applied mathematics that focuses on optimizing resource utilization, particularly in industrial timetabling.

In our daily life, timetabling presents a significant challenge for decision-makers due to the occurrence of numerous *unexpected events* during planning execution. However, most industrial timetabling models and solution approaches found in the literature assume a completely stable and deterministic environment, often neglecting uncertainties. As a result, implementing deterministic industrial timetabling solutions, even when they are theoretically optimal, remains challenging in real-world scenarios where various uncertainties can disrupt the decision-making process when dealing with a pre-generated schedule produced by an optimizer (e.g., CPLEX, Timefold). Among the existing works in the literature, a non-exhaustive list of potential disruptions in the context of industrial timetabling includes: (1) the arrival of a new high priority task [2], (2) the addition or cancellation of a task [3], (3) the unavailability of resources to be delivered [4], (4) the unavailability of a worker [5], (5) a change in task priority [2], (6) the variability of task processing time [6].

To address the timetabling rescheduling challenge, collaboration between humans and machines becomes necessary. When an optimal schedule is generated by an optimizer, unexpected events, such as an unexpected staff absence, may occur during execution. In such cases, a recommendation system can propose several alternative strategies to handle the disruption. By integrating human expertise, decision-makers can select the most appropriate strategy based on their business experience. The chosen strategy is then fed back into the optimizer, allowing for a rapid update of the schedule while accounting for the unexpected absence. In this paper, we present a recommendation system-based framework for updating an optimal schedule generated by a metaheuristic algorithm, subject to several constraints. The update process mainly relies on construction heuristic algorithms, which offer a time-efficient approach to obtaining an up-to-date, near-optimal solution with reduced computation time, as suggested in [7]. This approach avoids the need to launch a full optimization process during the solution exploration stage of the metaheuristic search process.

The remainder of this paper is structured as follows. Section 2 presents a synthesis of the literature on various rescheduling approaches applied to address the challenges of industrial timetabling rescheduling. Section 3 outlines the assumptions and research scope of this study. In Section 4, we present the proposed system and describe the implementation environment, which is based on the Timefold AI Solver. In Section 5, we detail the experiments of our framework, beginning with the description of the test use case and test environment, and ending with an illustration of our recommendation-based framework for rescheduling. Finally, Section 6 concludes this research and discusses potential future directions.

## 2 Previous works in industrial timetabling rescheduling

Rescheduling in industrial timetabling involves updating schedules to accommodate uncertainties that arise during the execution of a pre-generated schedule. The choice of suitable approaches plays a crucial role in determining the quality of the rescheduling solution. In [8], an event-driven approach is proposed to address the job shop problem in dynamic manufacturing system. The approach features an architecture that integrates the entire rescheduling process, from information acquisition and optimization to rescheduling. In [9], a decision making schema : Smart Scheduling, is presented to address unexpected and disruptive events in the manufacturing process during production planning generation. In Operations Research (OR) approximated method, simulated annealing is used in [10] to manage unexpected job arrivals in manufacturing. The objective is to determine a feasible rescheduling strategy that minimizes the maximum weighted tardiness costs while limiting disruptions to the original schedule. In [11], a reordering algorithm based on Mixed-Integer Linear Programming (MILP) is proposed to generate a feasible schedule and subsequently reallocate it optimally within the production program. The method incorporates the flexibility to reschedule jobs in response to changes in operational parameters. This approach is applicable to specific scheduling problems arising in the paper-converting and pharmaceutical industries. In [12], a hybrid multi-population genetic algorithm and constraint programming approach is used to solve real-time flexible production rescheduling. This method considers a dynamic flexible job shop environment while incorporating shop floor disruptions. In [13], a MILP-based reactive scheduling framework for short-term scheduling problems is developed to update the initial schedule and provide an immediate response to unexpected events, such as equipments breakdown. To enable a more flexible handling of Flexible Manufacturing Systems (FMS) in the generation of optimal schedules, the authors in [14] propose a soft computing approach. This method combines (1) schedule generation techniques integrating Petri nets, discrete-event simulation, and memetic algorithms; (2) a rescheduling environment that accounts for machine failures, with the objective of optimizing both makespan and total weighted tardiness.

Machine Learning (ML) has been introduced in several studies to address rescheduling challenges. In [15], a framework combining metaheuristic optimization algorithms and ML is proposed. This framework tackles the scheduling problem using a hybrid metaheuristic approach, and then applies a ML classification model to identify rescheduling patterns. The goal is to support rescheduling decisions when unexpected events occur in a flexible job-shop scheduling problem. Similarly, another ML-based method involves predicting a variable strongly correlated with the rescheduling requirement, using a comparable classification model [16]. Another category of ML approaches that has been extensively studied in the context of industrial scheduling and timetabling is Reinforcement Learning (RL). An advantage of RL methods, including approaches based on Q-Tables [17] and Deep Q-Networks [18], is that they require fewer historical instances for training. The model can be continuously reinforced and improved through interaction with the environment in applied scenarios. However, RL approaches rely on a schedule state simulator that integrates selected actions and computes the corresponding rewards. These methods are particularly well-adapted for dynamic scheduling, as demonstrated in [19], where actions are not fixed repair strategies, but decisions made by agents within an environment to dynamically construct the schedule.

As another approach explored in previous works, the Multi-Agent System (MAS) has been proposed by [20], addressing production disturbances and rescheduling affected operations in a highly dynamic, online manufacturing environment. In [21], a digital twin-based prediction approach is introduced to handle unexpected order arrivals and to enable efficient rescheduling in flexible production workshops. This approach defines rescheduling strategies based on assumptions about potential unexpected events, allowing for schedule updates to be made in advance of the occurrence of these events. In [22], a simulation-based approach is presented for the test, validation and benchmarking of rescheduling methods, along with an industrial application of the proposed methods.

## 3    Research scope

### 3.1    Problem description

Machines used in manufacturing become less effective over time, which can affect product quality and production efficiency. To avoid unexpected breakdowns and maintain high performance, regular maintenance is necessary. One important type of maintenance is Preventive Maintenance (PM), which is planned in advance to reduce the risk of equipment failure and to shorten repair times.

PM is a key part of maintenance management. It involves studying the conditions of machines to decide the best ways to prevent problems before they occur. This helps to extend the life of equipment and improves overall production reliability. As a result, the PM scheduling problem was developed to manage and organize PM tasks effectively [23]. This problem focuses on three main decisions:
- When each maintenance task should be started,
- Which technician should perform it,
- Whether the technician has the right skills to do the assigned task correctly.

This scheduling process aims to use resources efficiently while keeping equipment in good working condition. In [23], PM scheduling problem is classified as a type of resource allocation and scheduling problem.

### 3.2    Research context and assumptions

This paper addresses PM scheduling problem in the presence of unexpected events that may arise during daily operations. The study is based on a real industrial environment, where machines are distributed across multiple production lines within a factory managed by a company. Each machine is associated with a set of PM tasks, and a group of technicians employed by the company is available to perform these tasks.

Depending on the chosen planning horizon (daily, weekly or monthly), the factory manager must determine a near-optimal schedule that assigns available technicians to the PM tasks that need to be performed within the defined horizon. Each PM task is characterized by several attributes, including a unique identifier, the specialization required to perform the task, and a deadline specified by the manager. This deadline is typically determined by the type of PM contract linked to the machine. For example, some contracts may require a technician to respond and perform the PM task within 24 hours. Technicians who are available during the planning horizon are also described by specific characteristics. These include their ID, availability, specialization and maximum workload capacity per day and per week. The planning is generated using an optimizer embedded within an ERP system, similarly to [24]. The constraints considered in our planning model are described in **Table 1**. They are classified according to their severity into three categories: hard, medium and soft. Thus, the quality of a schedule generated by the optimizer is assessed using a three-level scoring system, expressed in the format [hard: -100, medium: -100, soft: -100], where 0 represents the best possible score for each category. Violations of hard constraints result in an infeasible solution and must be strictly avoided. In contrast, violations of medium constraints are tolerated, though minimizing such violations leads to a better-quality solution. Soft constraints are primarily associated with the optimization of solution quality: the optimizer aims to maximize the soft score, striving to bring it as close to 0 as possible. Consequently, a higher soft score indicates a higher-quality solution.

**Table 1.** Description of considered constraints

| Constraint | Severity | Description |
|---|---|---|
| Opening hours | Hard | All PM tasks must be assigned during opening hours |
| Staff unavailability | Hard | All PM tasks must be assigned to available staff |
| Specialization | Medium | Staff must be assigned to the type of tasks they are specialized in |
| Deadline | Medium | PM Tasks must be completed before their deadline |
| Workload limit | Medium | Staff must not work more than a fixed limit on weeks and days |
| Workload balance | Soft | Global workload must be balanced between all staffs |

As outlined earlier, the execution of a schedule on an operational day may be disrupted by unexpected events that impact the schedule initially generated by the optimizer. The first key assumption in our study is that *the schedule is prepared by an optimizer, subject to the constraints presented in* **Table 1** *in advance of the operational day, and that adjustments are subsequently made to accommodate any unexpected disruptions*. In this study, the types of unexpected events considered include the following:

- E1: The arrival of additional available staff during the operational day.
- E2: The unexpected absence of scheduled staff during the operational day.
- E3: The urgent need to perform additional PM tasks during the operational day.
- E4: The cancellation of previously scheduled PM tasks.

These events necessitate adjustments to the schedule to maintain operational efficiency and service quality.

The second assumption is that *adjustments to the pre-generated schedule are not made in real time, but through collaboration between the human (decision-makers) and the machine (recommendation framework)*. Our framework does not support automatic event detection through a digital twin network, and it cannot reschedule PM tasks immediately based on real-world changes. Instead, the framework is used before or during the execution of the original schedule, relying on iterative interaction between humans and the machine: when an event change occurs, the decision-maker is required to provide the updated information in the form of a disturbed schedule. In the case of E3, this could involve providing a schedule with some unassigned PM tasks. This schedule-update serves as the input to the framework to initiate the rescheduling process.

The final assumption *is related to the problem addressed in this paper, which is based on a real use case from one of our clients*. In this context, PM tasks are carried out by technicians within a single operational site. As a result, travel time between consecutive PM tasks is not considered, since movement within the site is quick and does not significantly affect the schedule. All technicians are assigned to the factories where the PM tasks take place. The constraints considered in this study, as listed in Table 1, were identified through several workshops conducted with the client, based on their specific requirements and operational specifications.

### 3.3 Research objectives and questions

To address the industrial timetabling rescheduling challenge for PM through human-machine collaboration, we propose a framework based on a recommendation system. A recommendation system is defined as a set of decision-support strategies designed to assist users in complex information environments to achieve specific objectives [25]. In our context, the objective is to *enhance human–machine interaction during the rescheduling of PM tasks in response to unexpected events*. Instead of letting the machine (the optimizer) generates solutions, we include the human expert in the process. Although optimization algorithms may produce mathematically optimal solutions in the field of Operations Research (OR), these solutions may not always align with the real needs or preferences of the business expert. This leads to the following research questions:

- RQ1: *How can we facilitate human-machine collaboration to deal with schedule disruptions efficiently?*
- RQ2: *How can we ensure a proper balance between quality and efficiency to fasten interactions while maintaining operator control over optimality?*

## 4 Recommendation system-based framework

### 4.1 Implementation environment

To ensure computational efficiency when generating rescheduling solutions, we selected Timefold[1] as the implementation environment for our framework. This choice has been guided by numerous reasons: (1) Timefold provides up to 10 built-in construction heuristics for solution generation as well as 7 local search-based metaheuristics to deeply improve found solutions. In OR, construction heuristics are known to be effective methods for producing high-quality approximate solutions, especially when computing time is limited [26]. Furthermore, local search, according to [27], forms the foundation of many meta-heuristic algorithms (e.g., Simulated Annealing [28]) for solving combinatorial optimization problems. (2) Timefold is open-source and developer-friendly, making it well-suited for building and adapting optimization models. This aligns with previous experiences using Timefold (formerly known as OptaPlanner) to develop optimization services [24], further supporting our decision. (3) Finally, Timefold offers additional support for rescheduling through the use of dynamic scheduling and construction heuristics, as discussed in the following sections. Together, these reasons make Timefold a solid foundation for building our recommendation system-based framework.

### 4.2 Focusing on construction heuristics in Timefold

Construction heuristics are used to initialize a solution by assigning planning values to variables defined within the planning entity. In the context of this study, the planning entity is the *PM task*, and the variables to be assigned are the *start time* and the *technician*. Thus, applying construction heuristics in our framework involves assigning a feasible start time and selecting an appropriate technician for each PM task.

---

[1] Timefold PlanningAI: https://timefold.ai/

Timefold construction heuristics module offers various strategies for solution initialization, which differ in how entity–value pairs are evaluated and selected. Three basic ***heuristic strategies*** are embedded in Timefold: (i) *entity-value combinations in a pool* (PO), (ii) *entities in a queue* (EQ) and (iii) *values in a queue* (VQ). The simplest heuristic is to put all *entity-value combinations in a pool* (PO) [29]. The combination resulting in the best score is retained, and all combinations for the attached entity are removed from the pool since it is now initialized. The preceding step is repeated until all entities have been assigned. A faster approach is to put *entities in a queue* (EQ) [30] and assign them one after the other. In this case, not all the entity-value combinations need to be considered at each step, which reduces the complexity and offers better scaling. However, the order in which entities are sorted in the queue matters. The third heuristic implemented in Timefold takes the opposite view to the second one and puts *values in a queue* (VQ) [30]. For each value, it is tested on all entities, and the most suitable entity is assigned to this value. This process continues until all entities have been assigned.

To improve efficiency, those strategies can be tuned in different ways. First, a parameter is dedicated to specifying the ***score behaviour*** on initialization, with two possible values: (i) *any* (AN) and (ii) *only down* (OD). The default value of this parameter is *any* (AN), which means that the score can change in any way upon initialization of a new variable. However, in the case where initializing a new variable can only worsen the score, it can be given the value *only down* (OD) specifying that a non-deteriorating move is a best move, which may greatly reduce the computation time by stopping early the search of entity-value combinations. Interestingly, another parameter is dedicated to tuning this stop early condition: the ***pick early type***. It can take 4 different values: (i) *never* (NE) is the general default value, which disable the early picking. (ii) When the score behaviour is set to *only down* (OD), the default value is *first non-deteriorating score* (ND) which stops when an assignment is found that does not worsen the score. (iii) To go further, options exist if execution speed is more important than quality: when set to *first feasible score* (FF), the search stops at the first solution with a non-negative hard score. (iv) In the case of *only down* (OD) behaviour, a last option is to set the *pick early type* to *first feasible score or non-deteriorating hard* (FN), which prevents searching for moves that improve an already negative hard score.

The early-stopping parameters discussed earlier highlight the importance of how entities and values are ordered in queues, since not all elements are guaranteed to be processed. To help sorting entities and values, 2 parameters have been proposed. The first one is the ***entity sorting*** *manner* which has two possible values: (i) *none* (EN), where entities are processed in the order they are encountered, and (ii) *decreasing difficulty* (ED), where entities are sorted based on a specified way. The reasoning behind this is that the more an entity is difficult to assign, the earlier it should be assigned, since assigning it late could result in important constraints violations. To be able to use this parameter, it is thus required to define a difficulty comparator for entities. For instance, in the case of PM tasks, the difficulty comparator can be based on tasks duration since longer tasks are less likely to fit in schedules at the end of the initializing process. Similarly, the second parameter is the ***value sorting*** *manner* which specifies how to sort values to try to fit in entity variables. 3 values are possible: (i) *none* (VN), (ii) *values are sorted by increasing strength* (VI) and (iii*) values are sorted by decreasing strength* (VD). The default option is VN. For VI and VD, the strength of a value is defined as how likely this value is to be assigned to an entity. VI is generally preferred because, once most entities have been assigned, it is easier to fit stronger values than weaker ones. However, in rare cases, using stronger values first (VD) can lead to better results.

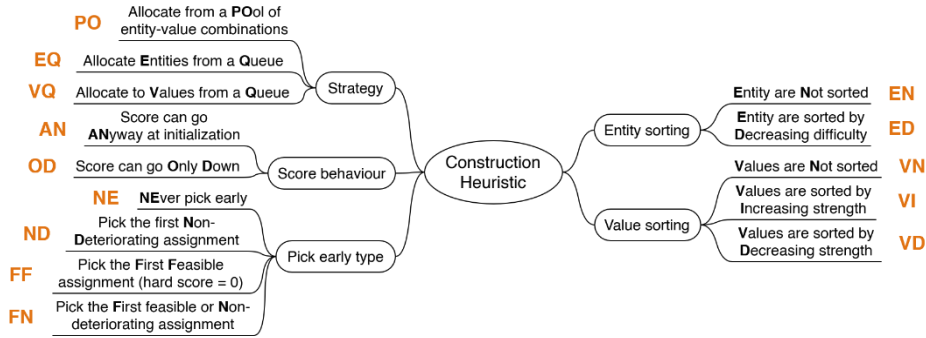

**Fig. 1.** Construction heuristics parameters overview

All the described parameters and their values can be retrieved in **Fig. 1** that provides an overview of this section. Abbreviations are used to lighten graphics in the experiments section. Those are highlighted in the figure below.

### 4.3 Recommendation system description

When unexpected events (see Section 3.2) occur, the previously optimized schedule may become suboptimal (E1 and E4) or some unassigned elements can appear, resulting in a partially initialized schedule (E2 and E3). Responding to RQ1, the proposed recommendation system aims at facilitating human-machine collaboration by offering different suggestions to help the user getting a new solution in response with its needs. Those strategies are:

- *Option 1 – Full recovery*: the full optimization process can be launched on the updated problem to ensure an optimal solution can be reached.
- *Option 2 – Manual assignments*: in cases E2 and E3, since the schedule is now partially initialized, unassigned PM tasks must be assigned. This can be done manually with a list of detailed suggestions. The solution may not be optimal but respect the user needs and reduce changes.
- *Option 3 – Automatic assignments*: this is the same as *Option 2* but instead of requiring a choice between multiple suggestions, the best suggestion (according to the induced score change) is automatically chosen. The solution may be suboptimal and unaligned with user needs, but the computation time and changes are controlled.
- *Option 4 – Dynamic scheduling*: in cases E1 and E4, since the schedule is still initialized, we just want to relaunch scheduling based on the previous solution. For stability, the user can pin some entities that must not change. This may prevent optimality from being achieved, but it gives more control to the user.

The scope of this paper is to deal with *Option 2* and *Option 3*. Interestingly, construction heuristics can be used on partially initialized solutions. This is how we can initialize unassigned elements automatically. Also, when dealing with single unassigned entities, it is possible to get all evaluated values with the induced score difference. The manual repair feature of our recommendation system is based on this principle. Treating unassigned entities one-by-one, assignment possibilities are shown to the user who can choose among them, relying on the constraints' violation changes descriptions. This also brings explainability to the automatic initialization feature, seeing why a particular assignment would be chosen based on the constraints and allowing the user to tune those constraints to get results better suited to the business needs.

To respond to RQ2, we aim at offering users different options that balance computational time and response quality by adjusting the parameters of construction heuristics. Therefore, Section 5 presents an experimental study which aims at selecting construction heuristics parameters according to problem size based on results quality and computation time.

### 4.4 Framework overview

Consider the following scenario: a problem instance is initially passed to a Timefold-based optimizer that applies (1) a construction heuristic for initializing the solution and (2) at least one local search phase to find an optimal solution. Supposing this is a satisfying solution, it can be given to a staff crew. At some point, an unexpected event occurs, assumed to be of kind E1 to E4 (see Section 4). Consequently, the schedule is updated and might no longer be optimal, or even partially initialized.

This is the entry point of our proposed framework, as illustrated by **Erreur ! Source du renvoi introuvable.**. Whatever the type of events disturbing the planning, the operator can interact with our recommendation system to choose an appropriate strategy to repair the schedule. If the schedule is no longer initialized, options 1 (*full recovery*), 2 (*manual assignments*) and 3 (*automatic assignment*) are proposed since applying a construction heuristic is required, else options 1 and 4 (*dynamic scheduling*) are proposed. A repaired solution is then built, depending on the chosen strategy, and suggested to the user. At this point, the user reviews the solution and can accept it or tune it manually, and interact again with the recommendation system until a satisfactory state is found.

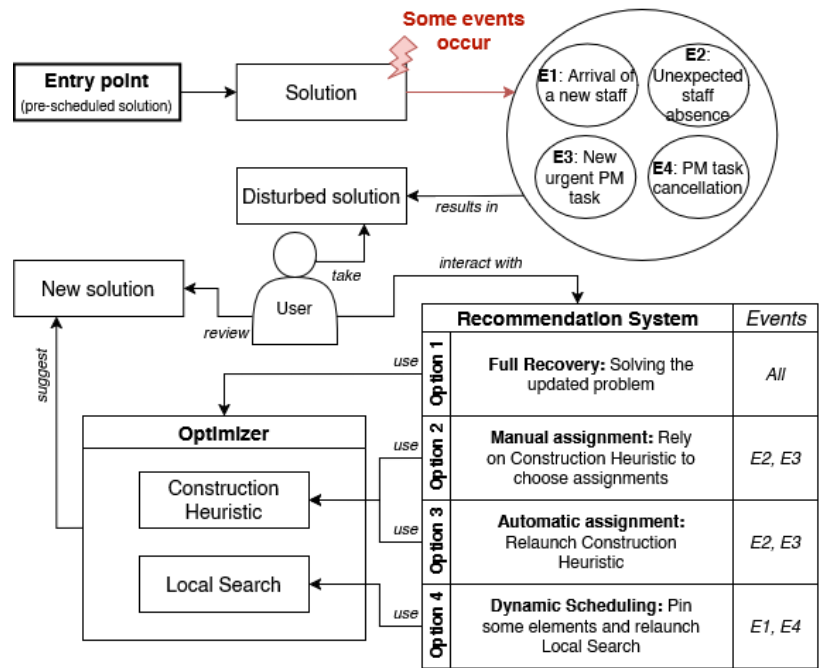

**Fig. 2.** Overview of our recommendation system-based framework

In this study, we focus on options 2 and 3 that rely on Timefold construction heuristics and offer more control to the user. Note that option 3 can be declined in multiple profiles if multiple construction heuristics configurations are given. For instance, if there is a very fast heuristic that results in poor score and a very slow one that results in good score, both can be proposed to the user for the automatic assignment. The following experiments section is therefore devoted to the study of trade-offs between quality and efficiency.

## 5 Experiments

This section presents an experimental study on Timefold-provided construction heuristics and a prototype for the proposed framework. The experimental study aims at comparing the construction heuristics parameters on solution quality and computational performance when the problem size varies. Note that construction heuristics are applied in almost all options of our recommendation system. The purpose of the experiment is to identify the most effective heuristic that provides the best balance between solution quality and computing time among all those available in Timefold.

### 5.1 Use case description

For the purpose of comparing construction heuristics on different problem sizes, a problem instance generator has been developed according to the use case described in Section 3.

The generated problems are described by numerous parameters at different levels. At experiment level, the following elements are defined: (1) the *working days and hours* for all staff members, which determine their availability and unavailability; (2) the *time granularity*, which is used to define time slots and possible values for assignments; (3) the *limit on working hours per days and weeks*, which applies to all staff and is used by the *workload limit* constraint.

The generator helps dealing with different scales of problems by defining parameters at scale level. These include: (1) the *range of durations* for generated PM tasks, expressed in number of slots based on the chosen time granularity; (2) *size of the horizon*, which defines the total number of possible assignment slots; (3) the *number of different specializations*, used by the specialization constraint; (4) the *PM tasks with deadlines rate*, which determines how many tasks must be scheduled before randomly generated dates in the second half of the planning horizon; and (5) the *unavailability rate for staff members*, which decreases the number of available assignment slots by introducing randomly generated half-day unavailabilities, reflecting realistic operational conditions.

Finally, at problem level are defined (1) the *number of PM tasks* to assign and (2) the *number of staff members* available to be assigned to those tasks.

For experimental purposes, we have defined three problem scales: small, medium, and large, each represented by three distinct instances. Each instance is labelled using the format *{scale}{number}* (e.g., S1 for the first instance of the small-scale problem). The difficulty of an instance is characterized using an occupancy rate metric, defined as the expected total number of time slots required by all tasks divided by the total number of available working slots, as described in Equation ( *1* ). Here, $N_t$ and $N_s$ denote the number of tasks and staff members respectively, $r_U$ the unavailability rate, $d_H$ the number of slots in the horizon (which depends on working day and hours, horizon size and time granularity) and $\overline{d_t}$ the expected duration of a task. A higher occupancy rate indicates a more difficult problem, as it reflects a greater saturation of staff availability within the given planning horizon relative to the total duration of the scheduled tasks.

$$r_O = \frac{N_t \overline{d_t}}{r_U d_H N_s}$$

( 1 )

Generator parameters for our nine problem instances and associated occupation rates are given in **Table 2**.

**Table 2.** Instance generator parameters for the use case dataset

| Parameter | Value | | | | | | | | |
|---|---|---|---|---|---|---|---|---|---|
| | **Experiment Level** | | | | | | | | |
| Working days and hours | Monday to Friday / 8 a.m. to 6 p.m. | | | | | | | | |
| Time granularity | 10 minutes | | | | | | | | |
| Limit on working hours | 7 hours a day / 35 hours a week | | | | | | | | |
| | **Scale Level** | | | | | | | | |
| | Small | | | Medium | | | Large | | |
| Range of task durations | 1-3 slots | | | 2-6 slots | | | 3-12 slots | | |
| Size of the horizon | 2 days | | | 1 week | | | 2 weeks | | |
| Number of specializations | 2 | | | 3 | | | 4 | | |
| Tasks with deadlines rate | 10% | | | 20% | | | 30% | | |
| Unavailability rate for staff members | 10% | | | 20% | | | 30% | | |
| | **Problem Level** | | | | | | | | |
| | S1 | S2 | S3 | M1 | M2 | M3 | L1 | L2 | L3 |
| Number of PM tasks | 50 | 100 | 150 | 200 | 300 | 400 | 600 | 1000 | 1200 |
| Number of staff members | 3 | 5 | 7 | 6 | 9 | 12 | 12 | 20 | 24 |
| Occupancy rate | 30% | 37% | 38% | 56% | 56% | 56% | 89% | 89% | 89% |

## 5.2 Experimental design

Our experiments are designed to compare all applicable combinations of construction heuristics parameters presented in Section 4.2.

In our scheduling model, two planning variables are considered for each PM task: the assigned technician and the time slot. Currently, this setup is incompatible with the VQ (putting values in a queue) strategy since it is not possible to configure how both variables will be considered, so only the PO (entities and values in a pool) and EQ (entities in a queue) will be considered. Also, the workload balance constraint considered in this study have a score that can improve when assigning new values, so all the heuristics will be configured with *score behaviour* as AN since OD is inappropriate. However, all hard constraints scores can only worsen, meaning that FF and FN *pick early types* give equivalent results but with less efficiency for FF, so it will not be considered. In conclusion, our experiments will compare heuristics configured with different combinations of parameters: 2 strategies (PO and EQ), 3 pick early types (NE, ND and FN), 2 entity sorting options (EN and ED) and 3 value sorting options (VN, VI and VD). This results in $2 \times 3 \times 2 \times 3 = 36$ heuristics configurations.

Entity sorting (EN and ED) and value sorting (VN, VI, and VD) require the definition of a difficulty comparator for entities and strength comparators for the possible values of the two planning variables. The *difficulty comparator* ranks PM tasks according to their expected assignment complexity, based on two criteria:

- Task duration: longer tasks are considered more difficult to schedule.
- Scheduling urgency: among tasks of equal duration, those with earlier due dates are considered more difficult to schedule.

The *strength comparators* evaluate the suitability of planning values based on how likely they are to be a good fit for a PM task:

- For the assigned technician variable, since all technicians have the same workload limits, a technician with fewer unavailabilities within the planning horizon is considered stronger.
- For the assigned time slot variable, slots at the beginning of working days are considered stronger. This is because they are more flexible, being suitable for any task whose duration does not exceed the length of a working day.

To assess the performance of the heuristics, two key evaluation metrics are employed:

- Solution quality represented by final three-level score: When comparing two heuristics on a given problem, priority is given to the hard score. If hard scores are equal, the medium score is used as the next criterion, followed by the soft score.
- Execution Time: This metric evaluates the efficiency of each heuristic based on its execution time. To ensure reliability and account for runtime variability, ***each heuristic configuration is executed 10 times independently***. The average execution time and the corresponding standard deviation are then reported for each configuration.

Due to the significant execution time required by certain heuristics on large problem instances (L1 to L3), a decision was made to exclude the least efficient configurations from further analysis. To identify which configurations are too slow to be considered, we evaluated all heuristic combinations on small problem instances (S1 to S3) and analysed the evolution of execution times, focusing specifically on the strategy and pick early type parameters, since other configuration options were found to have minimal impact on runtime. The problem scale is calculated using Equation ( 2 ), where $N_t$, $N_s$ and $d_H$ represent the number of PM tasks, the number of technicians, and the number of time slots in the planning horizon, respectively. For readability in figures, we present results using $\log_{10}(S_{problem})$.

$$S_{problem} = (N_s * d_H)^{N_t}$$

( 2 )

The results are presented in **Fig. 3**, which clearly show that the PO strategy combined with NE or ND pick early types result in significantly longer execution times compared to other configurations. As a result, these combinations are excluded from further experiments to improve evaluation efficiency. This reduction eliminates $1 \times 2 \times 2 \times 3 = 12$ configurations, leaving a total of $36 - 12 = 24$ heuristic configurations for the remaining tests.

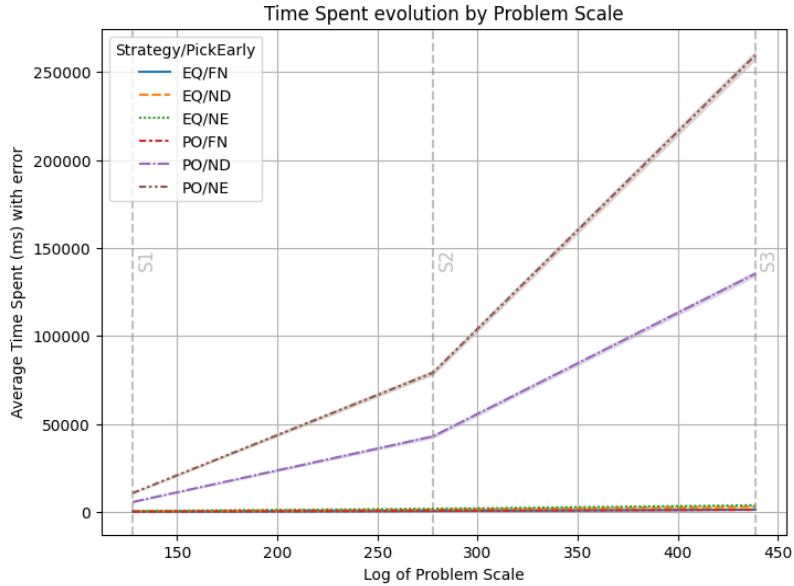

**Fig. 3.** Time spent evolution with error for strategy and pick early combinations, depending on problem scale

### 5.3 Numerical results

All results presented in this section are obtained using a computer with an 11th Gen Intel(R) Core(TM) i7-11800H @ 2.30GHz processor with 32.0Go of RAM, under the Windows 11 operating system. We first discuss results on score performances and then on time efficiency.

Since the hard score remains zero across all experiments, only the medium and soft scores are reported. Interestingly, we noticed than PO and EQ always gets the same scores for the FN pick early type, so we do not differentiate on the strategy when dealing with score performances. Each configuration is named according to its combination of options: for example, NE/EN/VN refers to the use of a strategy without pick early type, and no sorting for entities nor values. Results for the different problem instances are presented as bar charts, with separate graphs per problem. In each chart, the bar length represents the sum of the medium and soft scores, with the two components distinguished by different colours. Configurations are ordered from highest to lowest medium score. The resulting performance comparisons are shown in **Fig. 4**, organized by increasing problem difficulty across subfigures (a) to (i).

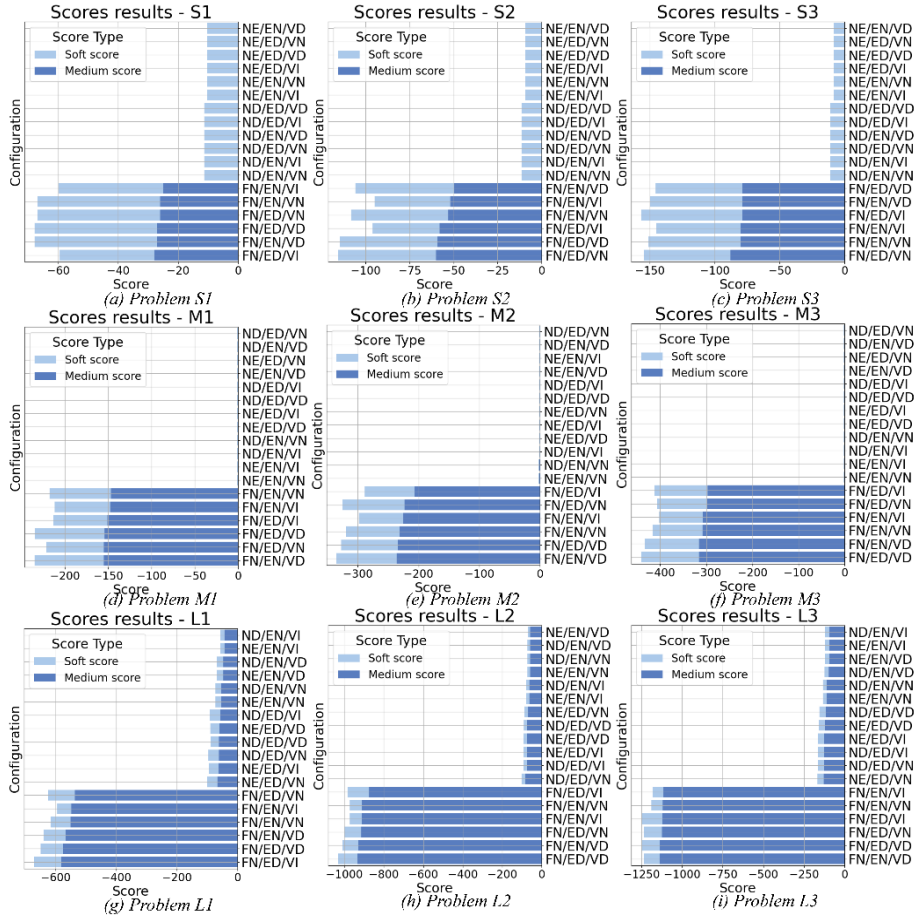

**Fig. 4.** Score results for tested configurations, sorted by worsening medium score

The results indicate a significant performance gap between the NE and ND pick early types on one side and the FN type on the other. As a reminder, the NE option never selects an assignment early, while the ND option chooses the first assignment that does not deteriorate the score. However, the FN option selects the first assignment that maintains or improves the hard score only, ignoring medium and soft scores. This behavior explains the poor performance observed with FN. Other configuration options appear to have a limited impact on the results, and the overall ranking of configurations varies significantly depending on the problem instance. Regarding sorting strategies, entity sorting (ED) tends to worsen performance or, at best, shows no measurable benefit. However, value sorting, whether by increasing (VI) or decreasing (VD) strength, frequently leads to improved solution quality.

We now evaluate the speed at which the different construction heuristics can initialize a solution. This aspect is particularly important in scenarios where a local search phase is applied afterward to further improve the solution. As in **Fig. 3**, the execution times are grouped by strategy and pick early type, with error margins displayed around each curve to illustrate standard deviation within each group. The results are presented in **Fig. 5**, showing how execution time evolves with increasing problem scale. For clarity, the logarithm of the problem scale is used, which explains the exponential trend observed in the curves. However, for larger problem instances (L1 to L3), heuristics using EQ strategy appear to exhibit a more linear growth in execution time.

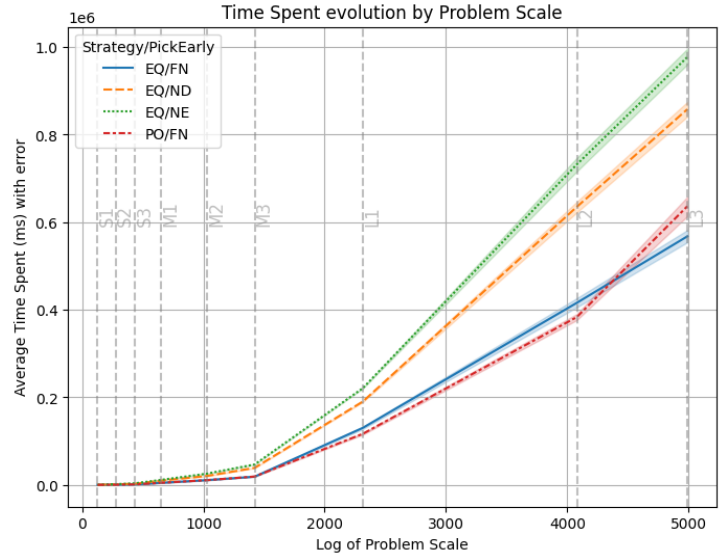

**Fig. 5.** Time spent evolution with error for strategy and pick early combinations, depending on problem scale

In summary, while never picking early (NE) and picking the first non-deteriorating assignment (ND) result in similar solution quality, the ND configuration offers significantly better execution speed, making it the more interesting choice overall. The third pick early strategy, FN, which considers only the non-deterioration of the hard score, provides an even greater speed advantage but at the cost of a substantial drop in solution quality. However, in some cases, computational time might be more important for a better user experience, so both ND and FN could be proposed to users within a recommendation system, depending on their performance–efficiency trade-off preferences. Finally, entity and value sorting have little impact on construction speed, and their effect on solution quality appears to be problem-dependent. Among our tests, sorting entities (ED) generally does not improve performance and should be avoided, whereas sorting values by increasing strength (VI) may lead to better outcomes and is worth considering.

### 5.4 Illustration of the recommendation system for rescheduling

To demonstrate the practical application of our approach, we present a snapshot of our recommendation system prototype. **Fig. 6** illustrates a simple scenario based on problem S1 in 7 steps. (1) The schedule has been initialized and (2) we add a new PM task. As a result, (3) the solution is no longer fully initialized, (4) so we trigger the recommendation system to respond. In this case, (5) the user selects the manual assignment mode (Option 2 of the system), which allows them to review a set of suggested assignments in detail and (6) validates. This enables the user to (7) make a choice based on their own preferences or business knowledge. The example highlights how the recommendation system-based framework facilitates cooperative decision-making, with the adaptability required to effectively handle rescheduling situations.

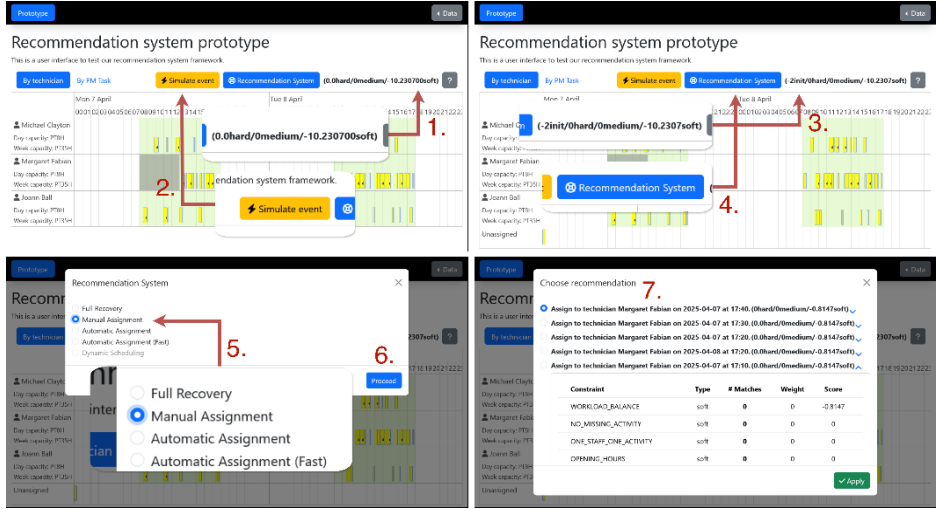

**Fig. 6.** Screenshots of our recommendation system prototype

## 6 Conclusion and research perspectives

In this paper, we present a recommendation system-based framework designed to enhance human-machine collaboration in industrial timetabling rescheduling. We apply this framework to address the rescheduling challenges associated with preventive maintenance. Our research begins with a literature review of existing approaches to industrial timetabling rescheduling. We then introduce the architecture of our recommendation system framework, which is built upon Timefold, a powerful AI-based planning engine used for implementation. In the experimental section, we evaluate nine instances derived from a real-world use case within our recommendation system, aiming to guide optimal decisions based on different construction heuristic configurations when unexpected events occur, and rescheduling becomes necessary. Finally, we demonstrate the complete process of our recommendation system using a simplified example (S1).

In the future, we plan to validate our framework through additional workshops with clients to assess how well it responds to RQ1. We are particularly interested in interactions between the user, the schedule and the recommendation system. We also aim to develop personalized rescheduling strategies corresponding to specific unexpected events, enabling users to apply these strategies directly within our framework. Lastly, we plan to explore a digital twin-based dynamic scheduling system. In this direction, unexpected events detected by a digital twin-enabled network would trigger real-time schedule adaptations, allowing the framework to automatically respond to disruptions as they occur and alert an operator on the fly.